\documentclass[%
 reprint,
amsmath,amssymb,
 aps,
]{revtex4-1}

\usepackage{graphicx}% Include figure files
\usepackage{dcolumn}% Align table columns on decimal point
\usepackage{bm}% bold math
\usepackage{epsfig,graphics,amsmath,amssymb}
\usepackage{epstopdf}
\usepackage{upgreek}
\usepackage{booktabs}
\usepackage{color}
\usepackage{draftwatermark}
\usepackage{pifont}
\usepackage{float}

\begin{document}
%\SetWatermarkText {preprint by Jiang}
%\SetWatermarkLightness{0.5}
%\SetWatermarkScale{3}
\preprint{APS/123-QED}

\title{Relative-intensity squeezing of high-order harmonic ``twin beams"}

\author{Shicheng Jiang$^{1}$ and Konstantin Dorfman$^{\dagger,1,2,3,4}$ }

\affiliation{1 State Key Laboratory of Precision Spectroscopy, East China Normal University, Shanghai , China}
\affiliation{2 Center for Theoretical Physics and School of Science, Hainan University, Haikou 570228,China.}
\affiliation{3 Collaborative Innovation Center of Extreme Optics, Shanxi University, Taiyuan, Shanxi 030006,China. }
\affiliation{4 Himalayan Institute for Advanced Study, Unit of Gopinath Seva Foundation, MIG 38, Avas Vikas,Rishikesh, Uttarakhand 249201, India.}

\email[Email:]{$^{\dagger}$ dorfmank@lps.ecnu.edu.cn }

\date{\today}% It is always \today, today,

\begin{abstract}
Relative intensity squeezing (RIS) is emerging as a promising technique for performing high-precision measurements beyond the shot-noise limit. A commonly used way to produce RIS in visible/IR range is generating correlated ``twin beams" through four-wave mixing  by driving atomic resonances with weak laser beams. Here, we propose an all-optical strong-field scheme to produce a series of relative-intensity squeezed high-order harmonic ``twin beams". Due to the nature of high harmonics generation the frequencies of the ``twin beams" can cover a broad range of photon energy. Our proposal paves the way for the development of nonclassical XUV sources and high precision spectroscopy tools in strong-field regime.
\end{abstract}

\maketitle
%\tableofcontents

The fundamental limit in high precision optical experiments is determined by the minimum uncertainty allowed by quantum mechanics, which is called ``shot-noise" limit (SNL).  Circumventing the SNL is crucial for the high precision optical experiments which had been demonstrated in e.g. gravitational wave detection \cite{Caves}. One of the possible solutions to circumvent SNL is to use squeezed states of light, where the noise in one quadrature is bellow the SNL with the expense of higher fluctuations in the other quadrature. Relative-intensity squeezing (RIS) of two correlated laser beams generated by four-wave mixing (FWM) has been demonstrated in numerous setups \cite{McCormick, Glorieux,MinXiao,jasperse} where the balanced homodyne differential measurement allows to observe signals beyond the SNL including the recent theoretical proposal \cite{zhenquan} and experimental demonstration of high-resolution FWM spectroscopy \cite{dorfmanpnas2021}.

The commonly used methods to generate nonclassical light source are based on low-order nonlinear effects in solids and gas ensembles \cite{tzallas1}. Recently, it has been demonstrated that the high-order harmonic generation in strong laser-matter interaction also shows quantum light features \cite{Gonoskov,Tsatrafyllis, cohen, javier,Tzallas2, lewenstein,Varro, Foldi}. Lewenstein et al. recently applied strong fields to generate ``Schr{\"o}dinger cat'' state \cite{lewenstein} and superposition of entangled and coherent states \cite{lewenstein2}. RIS is another possible quantum state of light which has yet to be demonstrated in strong field regime. 

The traditional FWM method to generate RIS in perturbation regime involves a strong pump ($\omega_{pu},{\bf k}_{pu}$) and a weak probe  ($\omega_{pr},{\bf k}_{pr}$) lasers, which give rise to a conjugate beam with frequency $\omega_c=2\omega_{pu}-\omega_{pr}$ and wave vector ${\bf k}_c=2{\bf k}_{pu}-{\bf k}_{pr}$. Finally, relative intensity squeezed state between the output probe and conjugate twin beams can be be measured by the homodyne intensity subtraction measurement. Due to the field-matter interaction is weak, FWM can be described by perturbation theory, where transitions driven by pump and probe lasers strongly depend on the bound states energies of the medium. Thus only two spectral components of relative-intensity squeezed beams are generated with specific frequencies in a double-Raman FWM process. We have recently shown that the three-step process in strong-field physics, e.g. tunneling ionization, acceleration by the laser and recombination, is analogous to stimulated Raman scattering \cite{PNAS,zheltikov}. We have further demonstrated that the strong-field induced polarization can be also expanded using semi-perturbative approach treating transitions between bound states perturbatively, while keeping interactions with continuum in strong-field approximation\cite{SFA}. This fundamental theoretical framework allows to extend many optical phenomena in perturbation regime to strong field regime. 

In this letter, we show that relative-intensity squeezing beween different pairs of harmonic fields can be also generated in all-optical strong laser regime. The proposed configuration is shown in Fig. 1. First, a strong infrared laser beam is split into two arms; one arm is used to interact with the medium in cell 1 to generate high harmonic field which is further sent to cell 2 acting as a seed probe, while the other beam acts as a strong pump pulse. 
\begin{figure}
    \includegraphics[width=3.5in,angle=0]{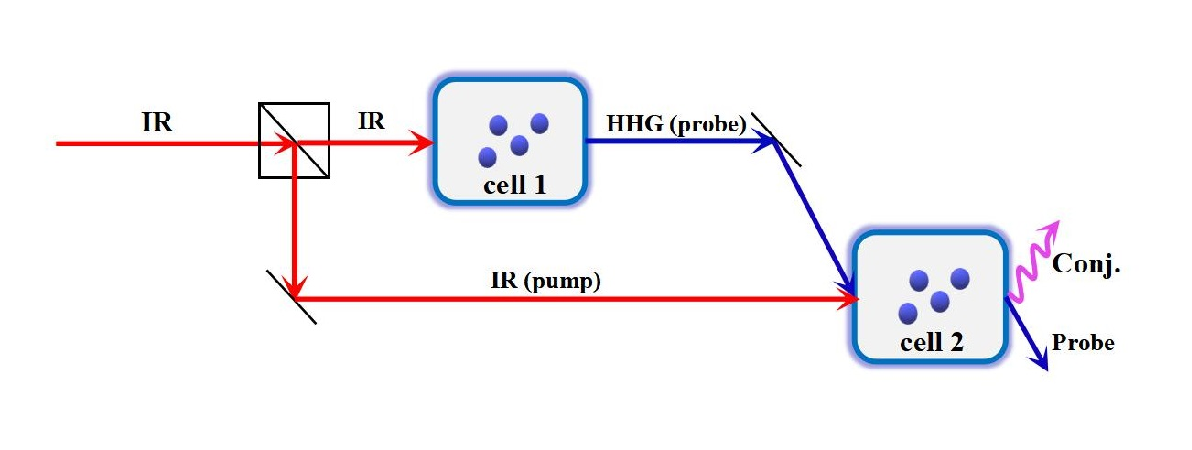}
    \caption{Schematic of the setup. A strong IR pump pulse is split into two arms, one of which is sent to cell 1 to generate high harmonic probe field in XUV region, the other beam is sent to mix with high harmonic probe pulse in cell 2. The output conjugate pulse is indicated by purple arrow. }\label{Fig_1}
\end{figure}

The interaction Hamiltonian between the system and the classical pump, quantized probe and output fields in cell 2, can be written in interaction picture with respect to the field Hamiltonian as 
\begin{equation}
H=H_{0}-q_e{\bf r}\cdot[\boldsymbol{E}_{p u}(t)+\sum_{\bf k} \hat{\boldsymbol{E}}_{\bf k}(t)]
\end{equation}
where $H_{0}$ is the field-free system Hamiltonian,$-q_e$ is the electron charge, $\boldsymbol{E}_{p u}(t)$ is the classical strong IR pump laser, $\hat{\boldsymbol{E}}_{\bf k}(t)$ contains the quantized incoming probe and emitted conjugate fields $\hat{\boldsymbol{E}}_{\bf k}=\sqrt{\frac{ \hbar \omega_{\bf k}}{\varepsilon_{0} V}}\left(\boldsymbol{\sigma} \hat{a} e^{-i \omega_{\bf k} t}+\boldsymbol{\sigma}^{*} \hat{a}^{\dagger} e^{i \omega_{\bf k} t}\right)$, $\boldsymbol{\sigma}$ is a unit vector of polarization, $\hat{a}$ and $\hat{a}^{\dagger}$ are the annihilation and creation operator respectively, $\varepsilon_{0}$ is  the vacuum permittivity , $V$ is effective interaction volume, and $\bf r$ is the dipole operator. Assuming a one dimensional case, 
\begin{footnotesize}
\begin{equation}
{r}=
\sum_{p, m}| p\rangle { d}_{p m}\langle m|+\sum_{p, m}| m\rangle { d}_{m p}\langle p |-\mathrm{i} \hbar\int d p|p\rangle \frac{ \partial}{\partial {p}}\langle p|.
\end{equation}
\end{footnotesize}
Here, $m$ and $p$ represent the bound and continuum states, respectively.  The third term is the continuum-continuum (c-c) transition operator which describes the accelertion of electron by strong laser. As we only focus on the above-threshold high harmonic signals, the dipole transitions between bound states are omitted here. It is more convenient to combine the field-free system Hamiltonian and the c-c coupling term as a new Hamiltonian $\mathcal{H}_0(t)=H_0-\mathrm{i}\hbar q_e \sum_{p, p^{\prime}}| p^{\prime}\rangle E_{pu}(t)\frac{\partial}{\partial p} \delta_{p p^{\prime}}\langle p|$.  For brevity, the remaining terms will be renamed as $\mathcal{H'}$. The total wavefunction (system+fields) can be expanded as
\begin{equation}
|\Psi(t)\rangle=\mathcal{U}_{0}\left(t, \tau_{0}\right) \exp _{+}\left[-\frac{i}{\hbar} \int_{t_{0}}^{t} d \tau \mathcal{H}_{I}^{\prime}\right]\left|\Psi\left(\tau_{0}\right)\right\rangle. 
\end{equation}
Here, we introduced the evolution operator $\mathcal{U}_{0}\left(t, \tau_{0}\right)$ with the $\mathcal{H}_0$, and $\mathcal{H}_{\mathrm{I}}^{\prime}(t) =\mathcal{U}_{0}^{\dagger}\left(t, t_{0}\right) \mathcal{H}^{\prime}(t) \mathcal{U}_{0}\left(t, t_{0}\right)$.  The subscript $ I$ stands for the``interaction representation" with respect to $\mathcal{H}_0$.  The notation ``exp$ _{+}$'' denotes the infinite expansion over $\mathcal{H}_{\mathrm{I}}^{\prime}(t) $\cite{shaulbook}. The total wavefunction reads
\begin{small}
\begin{align}
&|\Psi(t)\rangle=\mathcal{U}_{0}\left(t, \tau_{0}\right)\left|\Psi\left(\tau_{0}\right)\right\rangle+\sum_{n=1}^{\infty}\left(-\frac{i}{\hbar}\right)^{n} \int_{t_{0}}^{t} d \tau_{n}\cdots \int_{t_{0}}^{\tau_{2}} d \tau_{1} \nonumber\\
 &\times\mathcal{U}_{0}\left(t, \tau_{n}\right) \mathcal{H}^{\prime}\left(\tau_{n}\right) \mathcal{U}_{0}\left(\tau_{n}, \tau_{n-1}\right)\cdots \mathcal{H}^{\prime}\left(\tau_{1}\right) \mathcal{U}_{0}\left(\tau_{1}, \tau_{0}\right)\left|\Psi\left(\tau_{0}\right)\right\rangle.
\end{align}
\end{small}
This expansion is accurate and can reach convergency as long as the coupling term in the rough estimate $E_{pu}(t)d_{mp}<(E_m+U_p$), where $E_m$ and $U_p$ are the energy level of bound state $m$ and ponderomotive energy, respectively. We discuss the different order expansions in detail in the Supplementary Materials (SM). The XUV fields generated by HHG are typically weak, so the lowest order perturbative expansion with respect to the probe and conjugate fields is sufficient\cite{caowei}. As discussed in the SM, the lowest order expansion for each field along gives photon ionization by solely pump or probe fields. The second order expansions contain five components, one of which is equivalent to the well-known strong-field approximation model for HHG\cite{SFA}. For the third and fourth orders, the situation becomes more complex. While, following the standard approach\cite{loss,OE-FWM,agarwal}, the two-mode squeezing can be generated if the Hamiltonian has the form of $\xi \hat{a}^{\dagger} \hat{b}^{\dagger}-\xi^* \hat{a}\hat{b}$, where $\hat{a}^{\dagger}(\hat{a})$ and $\hat{b}^{\dagger}(\hat{b})$ are the creation (annihilation) operators for the two modes. Thus, the leading order which contributes to the RIS is the third order response.The corresponding wavefunction reads 

\begin{equation}
|\Psi(t)\rangle^{(3)}=|\Psi(t)\rangle_{a}^{(3)}+|\Psi(t)\rangle_{b}^{(3)},
\end{equation}
where 

\begin{align}
& |\Psi(t)\rangle_a^{(3)}=\zeta_a \hat{a}_c^{\dagger} \hat{a}_{p r}^{\dagger}\left|\Psi\left(\tau_0\right)\right\rangle, \\
& |\Psi(t)\rangle_b^{(3)}=\zeta_b \hat{a}_c^{\dagger} \hat{a}_{p r}^{\dagger}\left|\Psi\left(\tau_0\right)\right\rangle,
\end{align}
and 
\begin{small}
\begin{align}
\zeta_a & =\sum_{n, q} \frac{-i\sqrt{\omega_c^3 \omega_{p r}^3} \mu_{e g}^c\left(n \omega_{p u}\right) \mu_{g, e}^b  \delta\left(\omega_c+\omega_{p r}-n \omega_{p u}\right)}{\hbar^2\varepsilon_0 \pi^2 c^3\left(\frac{E_e}{\hbar}+\omega_{p r}-\frac{E_g}{\hbar}-n \omega_{p u}-i \gamma\right)}e^{-i\frac{E_g}{\hbar} t} \\
\zeta_b & =\sum_{n, q} \frac{-i\sqrt{\omega_c^3 \omega_{p r}^3} \mu_{e g}^c\left(n \omega_{p u}\right) \mu_{g, e}^b  \delta\left(\omega_c+\omega_{p r}-n \omega_{p u}\right)}{\hbar^2\varepsilon_0 \pi^2 c^3\left(\frac{E_e}{\hbar}+\omega_c-\frac{E_g}{\hbar}-n \omega_{p u}-i \gamma\right)}e^{-i\frac{E_g}{\hbar} t}
\end{align}
\end{small}
and $|\Psi(\tau_0)\rangle=|g\rangle\otimes|\eta\rangle\otimes\Pi_{c}|0\rangle_{c}$ is the initial total wave function. Here $g$ represents the ground electronic state, $\eta$ is the coherent input probe field and the subscripts ``c" and ``pr" denote the conjugate and probe fields, respectively. $E_{e(g)}$ is the field-free eigen energy for the excited state $e$ (ground state $g$), and $\omega_{pu}$ is the frequency of the pump laser. The electronic system, the incoming probe and the output conjugate fields are assumed initially in the ground state $|g\rangle$, the coherent state $|\eta\rangle$ and the vacuum state $|0\rangle$, respectively. $\mu_{mg}^c$ denotes effective transition dipole element describing the transition from bound state $|g\rangle$ to $|e\rangle$ which involves the ionization from the state $g$ to continuum state, free electron propagation and the recombination from the continuum to the bound state $e$ in the presence of strong IR. The third order wavefunction actually corresponds to the second order response for polarization. Similar to Ref. \cite{PNAS, OE}, the expression of the wavefunction can be read directly from the Feynman diagrams which are shown in Fig. S1 in SM. According to the diagrams and the explicit expressions for the third order expansion in SM, the physical picture corresponding to Eq. (6) and (7) is shown in Fig. 2. 

\begin{figure}[!htbp]
    \includegraphics[width=3.5in,angle=0]{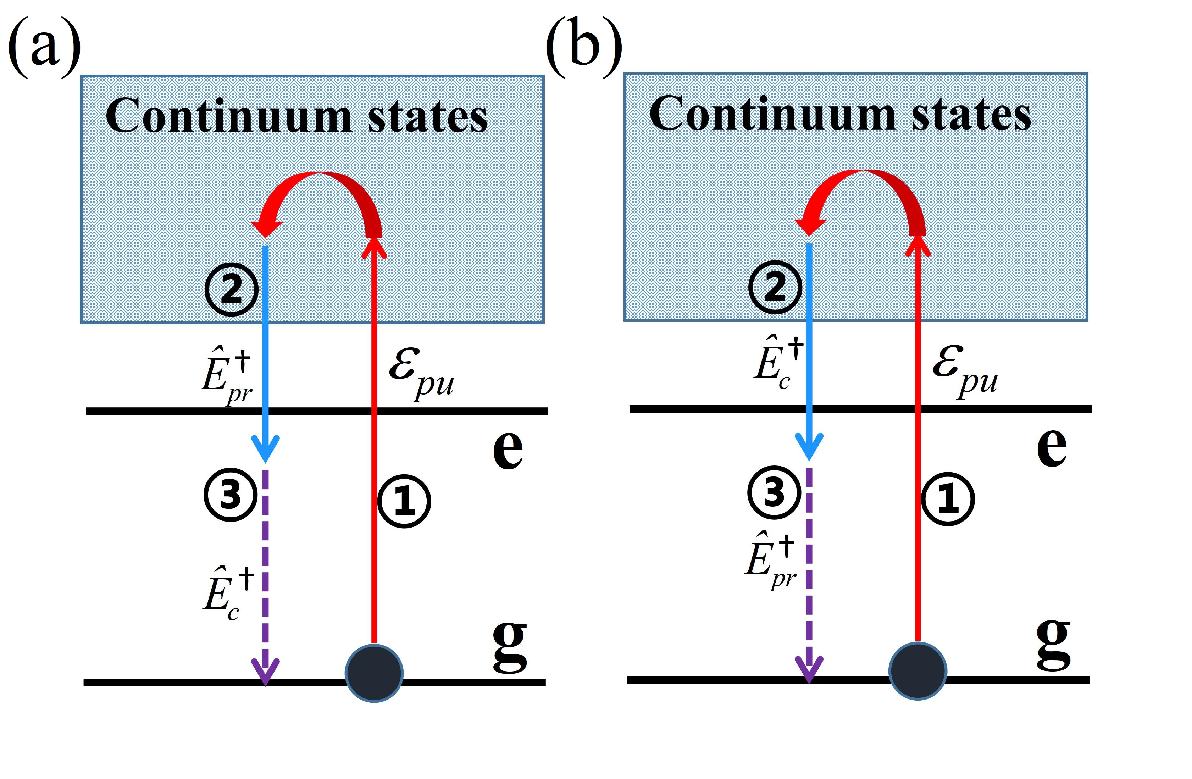}
    \caption{Energy level scheme and the transition paths corresponding to the polarizations for the conjugate (a) and the probe (b) fields.}\label{Fig_2}
\end{figure}

\begin{figure*}
\centering
    \includegraphics[width=6.5in,angle=0]{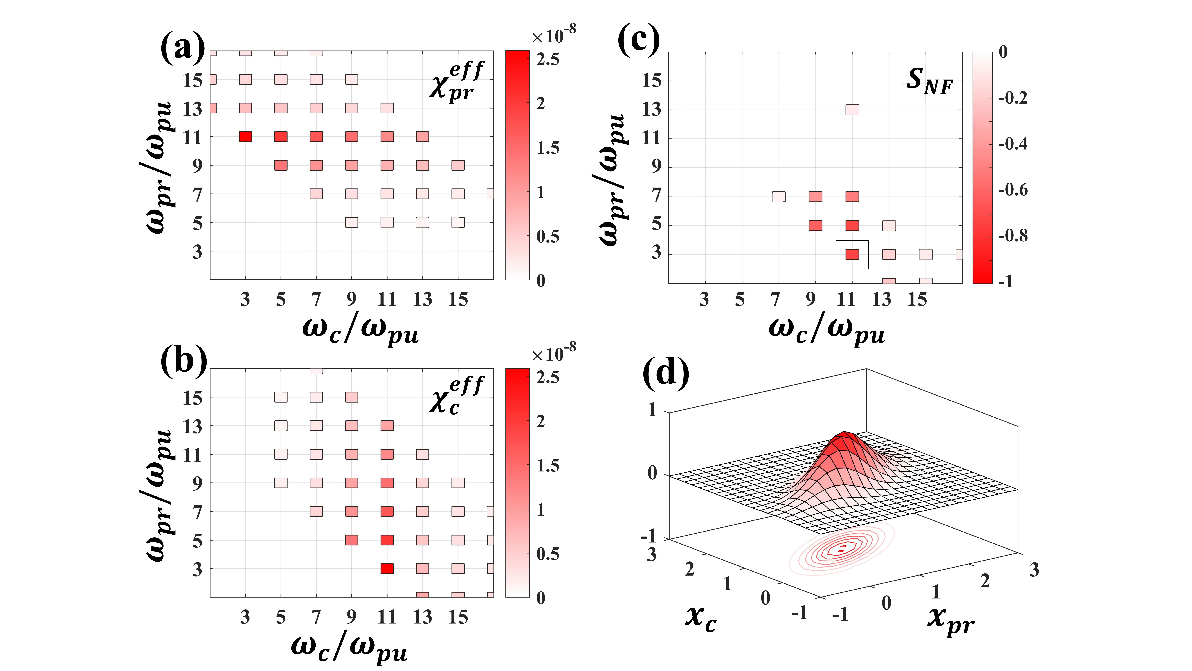}
    \caption{(a)-(b) The absolute values of $\chi_{pr}^{eff}$ and $\chi_{c}^{eff}$, respectively; (c) The noise figure in Eq. (17) vs harmonic orders of probe $\omega_{pr}/\omega_{pu}$ and conjugate $\omega_{c}/\omega_{pu}$; (d) The Wigner functions for the ``twin beams'' $(\omega_pr=3\omega_{pu},\omega_{c}=11\omega_{pu})$ in Eq. (19) vs $x_{pr}$ and $x_c$ with $p_c$ and $p_{pr}$ being set to be zero. The driving laser intensity is $5\times 10^{14} W/cm^2.$
}\label{Fig_3}
\end{figure*}
Eq. (6) describes a three-step process depicted in Fig. 2(a): \ding{172} The electron initially in the ground state $g$ tunnels directly into continuum; \ding{173} after acceleration in continuum by the strong pump laser, the ionized electron further recombines into the excited state $e$ of the target by stimulated transition induced by the high harmonic probe ($\omega_{pr},-{\bf k}_{pr}$) generated in cell 1; \ding{174} recombination into the ground state finally produces a wave-mixing phase-matched conjugate high harmonic field with frequency $\omega_c=n\omega_{pu}-\omega_{pr}$ and wavevector ${\bf k}_c=n{\bf k}_{pu}-{\bf k}_{pr}$, as indicated by the purple wavy line in Fig. 1. Fig. 2(b) corresponding to Eq. (7) is a concomitant by switching the time order between the probe and the conjugate fields\cite{GYYIN}. 

So far, we have treated the interaction between the system and light fields microscopically.  While, it is not enough to prove the effectiveness of our scheme. In the following, we consider a gaseous medium consisting of $N$ noninteracting atoms. Considering solely the third order response, the effective polarization operators for the output probe and conjugate fields are 
\begin{equation}
\begin{aligned}
& \hat{P}\left(\omega_{p r}\right)=\varepsilon_0 \chi_{p r}^{e f f} \hat{E}_c^{\dagger}\left(\omega_c\right) e^{i k_{n \omega_{p u}} z-i n \omega_{p u} t} \\
& \hat{P}\left(\omega_c\right)=\varepsilon_0 \chi_c^{e f f} \hat{E}_{p r}^{\dagger}\left(\omega_{p r}\right) e^{i k_{n \omega_{p u}} z-i n \omega_{p u} t}
\end{aligned}
\end{equation}
where$\hat{E}_{c / p r}^{\dagger}=\sqrt{\frac{\hbar \omega_{c / p r}^3}{\varepsilon_0 \pi^2 c^3}} \hat{a}_{c / p r}^{\dagger} e^{-i k_{c / p r} z+i \omega_{c / p r} t}$, $\chi^{eff}$ is an effective susceptibility describing high order responses of the medium, the explicit expression of which can be found in section 2 of the SM. $\chi^{eff}$ is proportional to the atom density $\rho=N/V$ inside the interaction volume\cite{Thiel,Bloembergen}:
\begin{equation}
\begin{aligned}
& \chi_{p r}^{e f f}=\left(-\frac{i}{\hbar}\right) \sum_{\omega_c}^{n\omega_{pu}>I_p} \frac{-\rho \hat{\mu}_{e g}\left(n \omega_{p u}\right) \hat{\mu}_{g, e}^b \delta_{\omega_{p r}, n \omega_{p u}-\omega_c}}{\left(\frac{E_e}{\hbar}+\omega_c-\frac{E_g}{\hbar}-n \omega_{p u}-i \gamma\right)} \\
& \chi_{c}^{e f f}=\left(-\frac{i}{\hbar}\right) \sum_{\omega_{p r}}^{n\omega_{pu}>I_p} \frac{-\rho \hat{\mu}_{e g}\left(n \omega_{p u}\right) \hat{\mu}_{g, e}^b \delta_{\omega_c, n \omega_{p u}-\omega_{p r}}}{\left(\frac{E_e}{\hbar}+\omega_{p r}-\frac{E_g}{\hbar}-n \omega_{p u}-i \gamma\right)} \\
&
\end{aligned}
\end{equation}
We then carry out the paraxial wave equation in the steady state (e.g. all time derivatives are zero) assuming slowly varying envelopes, perfect phase-matching and collinearity for the propagation. Additionally, if there is only one mode in the input probe field, simplified Maxwell equations read
\begin{equation}
\mathrm{i} \frac{\partial}{\partial z}\left[\begin{array}{c}
\hat{E}_{p r} \\
\hat{E}_{c 1}^{\dagger} \\
\vdots \\
\hat{E}_{c n}^{\dagger}
\end{array}\right]=H_{mxw}\left[\begin{array}{c}
\hat{E}_{p r} \\
\hat{E}_{c 1}^{\dagger} \\
\vdots \\
\hat{E}_{c n}^{\dagger}
\end{array}\right]
\end{equation}
where $E_{p r}^{\dagger}=\sqrt{\frac{\hbar \omega_{p r}^3}{\varepsilon_0 \pi^2 c^3}} \hat{a}_{p r}^{\dagger}, E_{c i}^{\dagger}\left(\omega_i\right)=\sqrt{\frac{\hbar \omega_{c i}^3}{\varepsilon_0 \pi^2 c^3}} \hat{a}_{c i}^{\dagger}$, the Hamiltonian-like matrix $H_{mxw}$ is 

\begin{equation}
\left[\begin{array}{cccc}
0 & i \kappa_{p r}\left(\omega_{c 1}\right) & \cdots & i \kappa_{p r}\left(\omega_{c N}\right) \\
i \kappa_{c 1}^* & 0 & \cdots & 0 \\
\vdots & \vdots & \ddots & \vdots \\
i \kappa_{c N}^* & 0 & \cdots & 0
\end{array}\right]
\end{equation}
with $\kappa_c=\frac{i k_c}{2} \chi_c^{e f f} \text { and } \kappa_{p r}=\frac{i k_{p r}}{2} \chi_{p r}^{e f f}$. By numerically solving the propagation equations above, finally we arrive at the input-output relationship for the probe and conjugate fields operators:
\begin{align}
\left\{\begin{array}{c}
\hat{E}_{p r}(z)=T_{1,1} \hat{E}_{p r}(0)+\sum_{j>1}^{N+1} T_{1, j}(z) \hat{E}_{c(j-1)}^{\dagger}(0) \\
\hat{E}_{c(k-1)}^{\dagger}(z)=T_{k, 1} \hat{E}_{p r}(0)+\sum_{j>1}^{N+1} T_{k, j}(z) \hat{E}_{c(j-1)}^{\dagger}(0) \\
\hat{E}_{p r}^{\dagger}(z)=T_{1,1}^* \hat{E}_{p r}^{\dagger}(0)+\sum_{j>1}^{N+1} T_{1, j}^*(z) \hat{E}_{c(j-1)}(0) \\
\hat{E}_{c(k-1)}(z)=T_{k, 1}^* \hat{E}_{p r}^{\dagger}(0)+\sum_{j>1}^{N+1} T_{k, j}^*(z) \hat{E}_{c(j-1)}(0)
\end{array}\right.
\end{align}
The parameter $z$ is the length of the one-dimensional cavity (the gase jet). The transformation matrix $T(z)=\mathrm{M(z)}^{-1} \mathrm{~K}(\mathrm{z}) \mathrm{M(z)}$, with $\mathrm{M}$ and $\mathrm{K(z)}$ being the eigenvector and eigenvalue matrices of the Hamiltonian ${H_{mxw}(z)}$, respectively. Fianlly, we obtain the variance for the relative-intensity as follows:
\begin{equation}
\begin{aligned}
 & \operatorname{Var}\left[\hat{\mathrm{I}}_{\mathrm{pr}}(\mathrm{z})-\hat{\mathrm{I}}_{\mathrm{ck}}(\mathrm{z})\right]= \\
&{\left[_{\mathrm{in}}\langle\hat{\mathrm{E}}_{\mathrm{pr}}^{\dagger}(\mathrm{z}) \hat{\mathrm{E}}_{\mathrm{pr}}(\mathrm{z}) \hat{\mathrm{E}}_{\mathrm{pr}}^{\dagger}(\mathrm{z}) \hat{\mathrm{E}}_{\mathrm{pr}}(\mathrm{z})\rangle_{\mathrm{in}}-{}_{\mathrm{in}}\langle\hat{\mathrm{E}}_{\mathrm{pr}}^{\dagger}(\mathrm{z}) \hat{\mathrm{E}}_{\mathrm{pr}}(\mathrm{z})\rangle_{\mathrm{in}}^2\right] } \\ 
+&{\left[_{\mathrm{in}}\langle\hat{\mathrm{E}}_{\mathrm{ck}}^{\dagger}(\mathrm{z}) \hat{\mathrm{E}}_{\mathrm{ck}}(\mathrm{z}) \hat{\mathrm{E}}_{\mathrm{ck}}^{\dagger}(\mathrm{z}) \hat{\mathrm{E}}_{\mathrm{ck}}(\mathrm{z})\rangle_{\mathrm{in}}-{}_{\mathrm{in}}\langle\hat{\mathrm{E}}_{\mathrm{ck}}^{\dagger}(\mathrm{z}) \hat{\mathrm{E}}_{\mathrm{ck}}(\mathrm{z})\rangle_{\mathrm{in}}^2\right] } \\ 
-&{}_{\mathrm{in}}\langle\hat{\mathrm{E}}_{\mathrm{pr}}^{\dagger}(\mathrm{z}) \hat{\mathrm{E}}_{\mathrm{pr}}(\mathrm{z}) \hat{\mathrm{E}}_{\mathrm{ck}}^{\dagger}(\mathrm{z}) \hat{\mathrm{E}}_{\mathrm{ck}}(\mathrm{z})\rangle_{\mathrm{in}}-{}_{\mathrm{in}}\langle\hat{\mathrm{E}}_{\mathrm{ck}}^{\dagger}(\mathrm{z}) \hat{\mathrm{E}}_{\mathrm{ck}}(\mathrm{z}) \hat{\mathrm{E}}_{\mathrm{pr}}^{\dagger}(\mathrm{z}) \hat{\mathrm{E}}_{\mathrm{pr}}(\mathrm{z})\rangle_{\mathrm{in}} \\ 
+& 2{}_{\mathrm{in}}\langle\hat{\mathrm{E}}_{\mathrm{pr}}^{\dagger}(\mathrm{z}) \hat{\mathrm{E}}_{\mathrm{pr}}(\mathrm{z})\rangle_{\mathrm{in}}{}_{\mathrm{in}}\langle\hat{\mathrm{E}}_{\mathrm{ck}}^{\dagger}(\mathrm{z}) \hat{\mathrm{E}}_{\mathrm{ck}}(\mathrm{z})\rangle_{\mathrm{in}}
\end{aligned}
\end{equation}
If the output probe and conjugate fields are in the coherent state (uncorrelated), then
\begin{equation}
\operatorname{Var}\left[\hat{\mathrm{I}}_{\mathrm{pr}}(\mathrm{z})-\hat{\mathrm{I}}_{\mathrm{ck}}(\mathrm{z})\right]_{SNL}={}_{\mathrm{in}}\langle\hat{E}_{p r}^{\dagger}(z) \hat{E}_{p r}(z)\rangle_{\text {in }}+{}_{\mathrm{in}}\langle\hat{E}_{ck}^{\dagger}(z) \hat{E}_{ck}(z)\rangle_{\text {in }}
\end{equation}
The noise figure which indicates the RIS degree between the probe (``pr") and conjugate (``ck") fields is defined as the ratio of the variance after squeezing to the variance in SNL 
\begin{equation}
S_{NF}= log_{10}\left[\frac{\operatorname{Var}\left(\hat{\mathrm{I}}_{\mathrm{pr}}(\mathrm{z})-\hat{\mathrm{I}}_{\mathrm{ck}}(\mathrm{z})\right)}{\operatorname{Var}\left(\hat{\mathrm{I}}_{\mathrm{pr}}(\mathrm{z})-\hat{\mathrm{I}}_{\mathrm{ck}}(\mathrm{z})\right)_{SNL}}\right]
\end{equation}
The case when $S_{NF}< 0 $ corresponds to the relative intensity squeezing, while the uncorrelated case yields $S_{NF}=0$.

Fig. 3(a-d) show the the effective susceptibilities, noise figure $S_{NF}$ and Wigner distribution  with $5.0\times 10^{14}$W/cm$^2$, 1240 nm pump laser. The pressure is assumed to be 0.5 bar at room temperature 300 K, and the propagation length z=2 mm. We successfully generate a series of relative-intensity squeezed ``twin beams" $(\omega_c, \omega_{pr})$. The dephasing rate $\gamma$ is set to be $1.2 \times 10^9 \mathrm{~s}^{-1}$\cite{dorfmanoe}. According to Eq. (11), the probe and conjugate fields satisfy the following energy conservation $\hbar\omega_c+\hbar\omega_{pr}=n\hbar\omega_{pu}$, which determines the shape of the $\chi_c^{eff}$ and $\chi_{pr}^{eff}$ in Fig. 3(a) and (b). Based on Eq.(15) and (16), the noise figure is inversely proportional to the sum of the photon numbers of probe and conjugate fields. For the parameters we used here, the gain of the probe field is small. Thus, the noise figure is approximately determined by the inverse of the effective suscetibility $\chi_c^{eff}$. Therefore, the noise figure in Fig. 3(c) has similar structure with Fig. 3(b) with the maximum squeezing degree reaching -1 db. 

As stated in \cite{scully}, the interface between classical and quantum light is elucidated by the photon statistics. For coherent (classical) state, the Wigner distribution \cite{wigner} has an isotropic Gaussion shape. The key feature of Wigner distribution for squeezed state is that the Gaussion distribution is narrowed in one quadrature and stretched in the other quadrature. For multi-mode squeezing case, the Wigner functions before ($W_i$) and after ($W_f$) the wave mixing  governed by the unitary transformation described by Eq. (14) are related by \cite{girish}
\begin{equation}
W_f\left(\alpha_{p r}, \beta_{c 1}, \cdots, \beta_{c N}\right)=W_i\left(\tilde{\alpha}_{p r}, \tilde{\beta}_{c 1}, \cdots, \tilde{\beta}_{c N}\right)
\end{equation}
where $\alpha_{p r}=\frac{\left(x_{p r}+i p_{p r}\right)}{\sqrt{2}}, \beta_{c j}=\frac{\left(x_{c j}+i p_{c j}\right)}{\sqrt{2}}$. The $\left(\tilde{\alpha}_{p r}, \tilde{\beta}_{c 1}, \cdots, \tilde{\beta}_{c N}\right)$ and $\left(\alpha_{p r}, \beta_{c 1}, \cdots, \beta_{c N}\right)$ are related by the unitary transformation given by Eq. (S37) in the SM. By using the input-output relationship, we obtain the Wigner distribution of the output radiation:
\begin{equation}
\begin{aligned}
&W_f\left(\alpha_{pr}, \beta_{c 1}, \cdots, \beta_{c N}\right)\propto  e^{-2\left|T_{1,1} \alpha+\sum_{m>1}^{N+1} T_{1, m} \beta_{c(m-1)}^{\dagger}-\eta\right|^2} \\
&\times e^{-2 \sum_j^N\left|T_{j+1,1}^* \alpha^*+\sum_{m>1}^{N+1} T_{j+1, m}^* \beta_{c(m-1)}\right|^2} 
\end{aligned}
\end{equation}
In Fig. 3(d), we plot the Wigner distribution of $W_f\left(\alpha_{p r},\beta_{c}\right)$ for the pair of ($\omega_{pr}=3\omega_{pu},\omega_c=11\omega_{pu}$) with other quadratures being set to zero. The narrowed shape of Wigner function indicates the correlation between the probe and conjugate fields via the relative intensity.

\begin{figure}[H]
    \includegraphics[width=3.5in,angle=0]{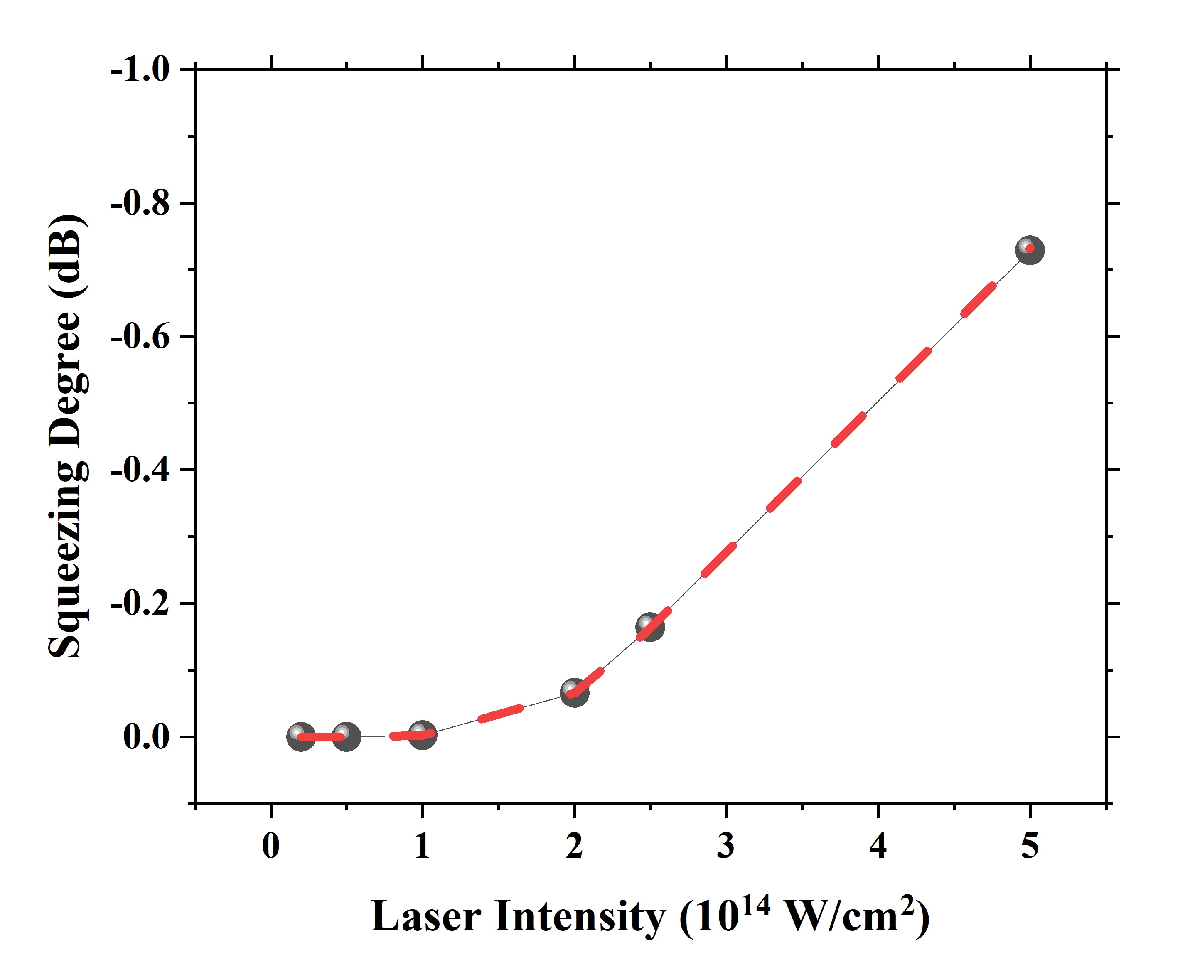}
    \caption{The dependence of noise figure for the photon pair $\left(\omega_{p r}=3 \omega_{p u}, \omega_c=11 \omega_{p u}\right)$ on the pump laser intensity. The filled cricles and red dashed line correspond to the multi-mode regime (Eq.15, 16) and the minimum coupling two-mode regime (Eq. 20, 21), respectively. }\label{Fig_4}
\end{figure}

Fig. 4 depicts the dependence of relative intensity squeezing noise on the pump laser intensity. The squeezing degree increases with the driving laser intensity. To gain insight into the scaling law, the limitation of minimum coupling between the probe and conjugate fields can be applied. In the minimum coupling limit, only single mode of probe and conjugate fields are included in the Maxwell equation, which yields 
\begin{equation}
\begin{aligned}
&\operatorname{Var}\left[\hat{\mathrm{I}}_{\mathrm{pr}}(\mathrm{z})-\hat{\mathrm{I}}_{\mathrm{ck}}(\mathrm{z})\right]_{S N L}^{min}=|g \sinh (\zeta)|^2+\left|g^{-1} \sinh (\zeta)\right|^2\\
&+\left(|\cosh (\zeta)|^2+\left|g^{-1} \sinh (\zeta)\right|^2\right) N_{p r}
\end{aligned}
\end{equation}
\begin{equation}
\begin{aligned}
\begin{aligned}
&\operatorname{Var}\left[\hat{\mathrm{I}}_{\mathrm{pr}}(\mathrm{z})-\hat{\mathrm{I}}_{\mathrm{ck}}(\mathrm{z})\right]^{min} =[|\cosh (\zeta)|^4+ \left\{|g|^2-|g|^{-2}\right\}\\
&|\cosh (\zeta) \sinh (\zeta)|^2-\cosh ^2(\zeta) \sinh ^2\left(\zeta^*\right)-\cosh ^2\left(\zeta^*\right) \sinh ^2(\zeta)\\
&+\left|g^{-1} \sinh (\zeta)\right|^4] N_{p r}+\left\{|g|^2+|g|^{-2}\right\}|\cosh (\zeta) \sinh (\zeta)|^2 \\
& -\cosh ^2(\zeta) \sinh ^2\left(\zeta^*\right)-\cosh ^2\left(\zeta^*\right) \sinh ^2(\zeta)
\end{aligned}
\end{aligned}
\end{equation}

where $g \equiv \sqrt{\frac{\kappa_{p r}}{\kappa_{ck}^* }}$, $\zeta \equiv z \sqrt{\kappa_{ck}^* \kappa_{p r}}$. The red dashed line in Fig. 4 gives the result predicted by the analytic expression in the minimum coupling limit. It is seen that in the regime of the present parameters, the minimum coupling limit works very well.

The key preconditions for generating squeezed twin-beams are 1) stimulated transition to the excited state in step \ding{173}; 2) recombination in step \ding{174} with the conjugate photon emitted. We noticed that survival of excited states in strong field has been widely investigated \cite{sandner, CDL, Jingchen}, which provide strong foundation to realize the proposal presented in this paper in the upcoming experiments. In section 4 of SM, we present the calculated noise figure for different pump laser wavelengths. It is obvious that the RIS is a general phenomenon, and not a specific consequence of the choice of laser parameters. In fact, if only a single strong pump laser is used, the generated harmonic modes can be still correlated\cite{lewenstein2}. We can further expect the emergence of RIS between different harmonic modes, while it would be impossible to separate the third-order process generating RIS photons from the dominant first-order process responsible for the standard classical HHG.

 Taking the advantages of quantum nature of optical signal in the strong field physics is a promising route to extend the applications of ultrafast science. For example, Ref. \cite{chekhova} has shown that the photon-number fluctuations can enhance the multi-photon ionization probability. One can further expect that high harmonic intensity could be enhanced by photon-number fluctuations in the driving laser. It also deserves exploring whether the resolution in attosecond experiments can be improved by the quantum nature of HHG \cite{worneroe}. More advantages of quantum light have been concluded in Ref. \cite{dorfmanRMP}. In summary, we demonstrated a possibility of all-optical method to generate multiple pairs of RIS ``twin beams" in XUV regime using multi-harmonic wave mixing in strong field-matter interactions. The Wigner distribution shows that the output fields are strongly correlated and show quantum squeezing properties in the photon statistics. Our proposal is a potential way to develop a new class of nonclassical light sources in XUV regime.

We thank Zhangjie Gao and Junjie Chen for discussion. We acknowledge the support of National Natural Science Foundation of China (12074124); Zijiang Endowed Young Scholar Fund, ECNU; Overseas Expertise Introduction Project (B12024).

\end{document}